\documentclass[angeo]{copernicus}
\newcommand{\farcm}{\mbox{\ensuremath{.\mkern-4mu^\prime}}}

\begin{document}

\l

\title{Errors in Scale Values for Magnetic Elements for Helsinki}
\runningtitle{Errors in Scale Values}
\runningauthor{Svalgaard}

\author[1]{Leif Svalgaard}
\affil[1]{Stanford University, Cypress Hall C3, 466 Via Ortega, Stanford, CA 94305, USA}
\correspondence{Leif Svalgaard (leif@leif.org)}


\received{31 October 2013}
\revised{}
\accepted{}
\published{}

\maketitle 

\abstract{
Using several lines of evidence we show that the scale values of the geomagnetic variometers operating in Helsinki in the 19th century were not constant throughout the years of operation 1844-1897. Specifically, the adopted scale value of the Horizontal Force variometer appears to be too low by $\sim$30\% during the years 1866-1874.5 and the adopted scale value of the Declination variometer appears to be too low by a factor of $\sim$2 during the interval 1885.8-1887.5. Reconstructing the Heliospheric Magnetic Field strength from geomagnetic data has reached a stage where a reliable reconstruction is possible using even just a single geomagnetic data set of hourly or daily values. Before such reconstructions can be accepted as reliable, the underlying data must be calibrated correctly. It is thus mandatory that the Helsinki data be corrected. Such correction has been satisfactorily carried out and the HMF strength is now well constrained back to 1845.
}

\introduction[Introduction and Rationale]  
After more than a decade of vigorous research \citep[e.g.][]{lea99, sea03, sva05, sva10, loc11} the magnitude of the Heliospheric Magnetic Field (HMF) near the Earth is well constrained from 1883 (probably even from 1872) to the present during which period sufficient and accurate geomagnetic data is available for calculation of the IDV-index \citep{sva05, sva10} that serves as a proxy for the HMF strength, $B$. There is still a healthy debate about the reconstruction before 1883 when geomagnetic data becomes sparse and subject to errors, especially in the more difficult to measure Horizontal Force.  

Although the IDV-index is calculated from the unsigned difference between the Horizontal Force at consecutive local midnight hours, it was already pointed out by \citet{sva03, sva05} and \citet[Figure 6]{sva10} that the IDV-index can be computed for any hour and for any geomagnetic element. Conforming with that stipulation, Lockwood and colleagues \citep[e.g.][]{lea13a,lea13b,loc13} have suggested to reduce the influence of noise in the early 19th century geomagnetic data by computing the average of the 24 individual time series of IDV calculated for each of the 24 hours of the day, dubbed IDV(1d). Although this procedure introduces unwanted variance because of the day-to-day variability of the [semi-regular] diurnal variation of the geomagnetic field, the `IDV-signature' is robust enough such as to reduce this extra variance to a second-order effect.  

Nevanlinna and colleagues \citep{nea92,nev04} have compiled archived geomagnetic observations from the Helsinki [IAGA designation HLS] magnetic observatory comprising over 2 million observations of H- and D-components measured during 1844-1912 with time resolution of 10 minutes to 1 hour. Because of disturbances from nearby electric tram lines and general curtailment of the observational program, reliable and complete daily records of hourly values are only available up through 1897. \citet{lea13a,lea13b} used this HLS data set to calculate IDV(1d). 

The validity of the resulting IDV(1d) series and of conclusions drawn from it, obviously hinges on the data being correctly calibrated. In this note we shall show that the scale value for the Horizontal Force is seriously incorrect [too small] during the interval 1866.0-1874.5 and that the scale value for the Declination is also incorrect [too small] during 1885.8-1887.5. These errors must be corrected before further inferences are drawn from the Helsinki data and before the results can be compared with and integrated into other reconstructions, in order to extend to earlier times the undisputed consensus reached for HMF B after 1882.

\section{The Summed Ranges}
\label{Summed-Ranges}
\citet{bar25,bar32} defined the u-measure as the average [over from 1 to 12 months] unsigned difference between the daily means of the Horizontal Force, formally equivalent to the proposed IDV(1d) index and used by \citet{sva05} in the derivation of their IDV-index. Before 1872 there were no readily available daily mean data for any magnetic observatory, so Bartels -- "more for illustration than for actual use" -- turned to use the `Summed Ranges' [designated $s$] supplied by \citet{moo10} as the main contributor to a proxy for the u-measure.  

\begin{figure*}[t]
\vspace*{2mm}
\begin{center}
\includegraphics[width=8.3cm]{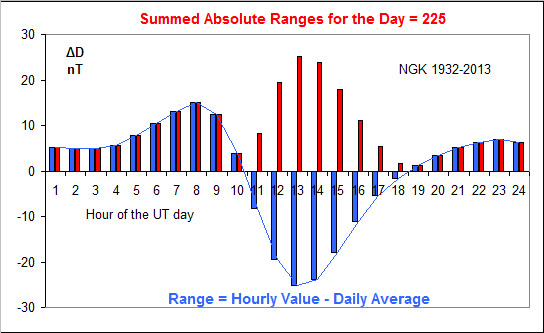}
\end{center}
\caption{Average diurnal variation of Declination (expressed in force units, nT) at Niemegk. On any given day, the variation consists of a pattern as shown here [although varying a bit from day to day] with superposed ‘noise’ from geomagnetic activity, thus increasing the variance; this increase is what we are interested in. The signed deviations [blue bars determined every hour – either from an instantaneous value on the hour or from the hourly mean] from the daily mean are converted to unsigned departures [red bars] which are then summed over the day giving [as Moos expressed it] the Summed Ranges for each day, denoted by $s$.}
\label{summed-ranges-def}
\end{figure*}

Bartels' interpretation of Moos' procedure and data \citep[Table 261]{moo10}  was: ``$s$ is derived from the mean diurnal variation of H at Bombay for each single month, expressed in departures from the average, and is the sum of these departures, summed without regard to sign". Using monthly means attenuates the irregular strong disturbances associated with the IDV-signature and gives undue weight to the regular daily variation, thus downplaying the role of true 'disturbances'. Moos was aware of this and on his effort of making a list of days classified as quiet or disturbed (ibid. page 421) remarked: ``[for] a list of the kind … involving a large personal equation, some additional data are clearly essential in order to make the classification more mathematically definite. The daily range, or preferably the summed ranges, figures of the diurnal inequality of \textit{each day}\footnote{our emphasis} would probably serve as the most appropriate data for this purpose; but as this is not possible on account of the heavy labour involved in their derivation (...he resorted to the monthly means eventually used by Bartels)" Here we shall build on that intuition (based on Moos’ extensive knowledge of the phenomenon) as we are no longer limited by computational power, provided data in digital tabular format is available. To make things explicit, Figure~\ref{summed-ranges-def} illustrates our interpretation of Moos’ prescription, emphasizing that by $s$ we shall henceforth in this paper mean $s$ derived from daily departures.

\subsection{Calculating IDV from Summed Ranges}
\begin{figure*}[b]
\vspace*{2mm}
\begin{center}
\includegraphics[width=16.6cm]{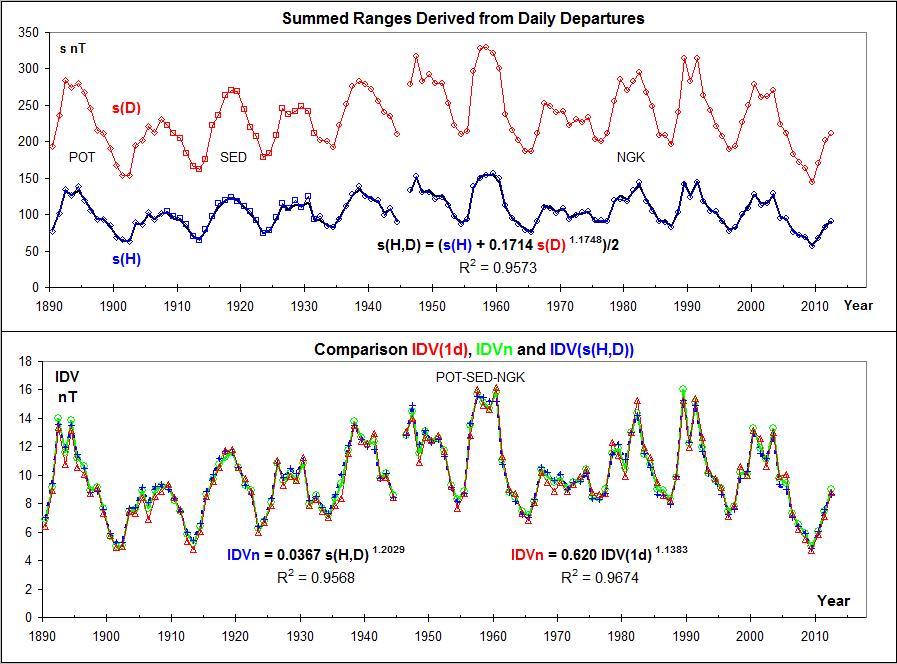}
\end{center}
\caption{(Upper panel) Summed Ranges derived from daily departures for Declination s(D) [red curve] and Horizontal Force s(H) [blue curve] for the combined POT-SED-NGK series. Each station's yearly value is marked with a different symbol [POT diamond, SED square, NGK circle]. The break in 1945 was caused by interruptions stemming from the Battle for Berlin during the final phase of WWII. The composite s(H,D) is added over s(H) as a black line. It is difficult to distinguish between the blue and the black lines. It is rare in this business to find such close agreement.\\
(Lower panel) IDVn (from mid\textit{\textbf{n}}ight values) for POT-SED-NGK [green line] compared to IDV computed from s(H,D) [blue dashed line]. Because the two curves are so close to at times be indistinguishable, each yearly value is also marked with a symbol: green circle for IDVn and blue plus sign for IDV(s(H,D)). Finally, the red curve and red triangles show IDV(1d) scaled to IDVn as indicated. We need that scaling because IDV(1d) is about 11\% higher than IDVn due to the day-to-day variability of the regular daily variation.}
\label{IDV-from-s}
\end{figure*}

We begin by calculating $s$ for both the Declination, $s(D)$ and for the Horizontal Force $s(H)$ for the German station Potsdam (POT, 1890-1907) and its replacement stations Seddin (SED, 1908-1931) and Niemegk (NGK, 1932-2012). Geomagnetic conditions were essentially the same at all three stations, because they were carefully placed with that in mind, so we can treat the data as homogeneous from a single station, Figure~\ref{IDV-from-s}.

By inspection it is clear that s(D) and s(H) are highly correlated, in fact: $s(H) = 0.1714s(D)^{1.1738}$ (The coefficient of determination R$^{2}$= 0.9573 is calculated from the linear fit of the logarithms), see also Figure \ref{Regression-sH-sD} in the SI. We can then form the average $s(H,D)$ of observed s(H) and s(H) calculated from s(D) as shown by the black line in the upper panel of Figure~\ref{IDV-from-s}. We can calculate the IDV-index for this particular station chain the usual way using unsigned differences between the hour following local solar midnight, call it IDVn. The series of IDVn and s(H,D) are also highly correlated:  $IDVn = 0.0367s(H,D)^{1.2029}$ (R$^{2}$= 0.9568); Figure \ref{Regression-IDV-s} in the SI. It is rare to find correlations that significant. The conclusion is that given either s(D) or s(H) or both, we can calculate a very close approximation (blue curve) to the usual IDVn (green curve) as shown in the lower panel of Figure~\ref{IDV-from-s}. This is particularly important for early stations where H is often very noisy, while D is well-observed (or at times even the only component observed).

\subsection{Calculating IDV from IDV(1d) or u-measure}
Because the u-measure was found to be a good proxy for IDVn \citep[e.g.][]{sva05} we expect the equivalent IDV(1d) index \citep{lea13a} to be as well. The red line in the lower panel of Figure~\ref{IDV-from-s} bears this out:  $IDVn = 0.620IDV(1d)^{1.1383}$ (R$^{2}$= 0.9674) for this homogeneous dataset, consistent with the finding by \citet[page 13]{may80} ``the u index … certainly suffers from intrinsic defects … One might suspect a contamination by the regular variation, since its day-to-day variability should contribute to the interdiurnal variability. However, we tried to evaluate the importance of this contamination and were astonished at its relative smallness." So, we have essentially three different ways of estimating IDV. This also holds for other long-term homogeneous station sets, e.g. PSM(Parc Saint-Maur)-VLJ(Val Joyeux)-CLF(Chambon-la-For\^et). As long as we limit ourselves to stations far enough ($>10\degree$) from the auroral zone these three different methods yield comparable and highly correlated values.

\section{Scale Values for Helsinki Data}
\label{Scale-Values}
The original archived data for Helsinki Observatory (situated at 60\degree 10\farcm{4}N, 24\degree 56\farcm{9}E) was given in `scale units' which must be converted to force units [nT, called gammas in older literature] or angles [typically tenths of arc minutes]. The scale units must be converted into physical units. The usual scheme calculates the physical values from the scale units like this: $phys.~value = base~value + scale~value * (scale~units + instrument~corrections)$.  Often the $base~value$ and the $instrument~corrections$ are not known and the magnetometer can be characterized only as a `variometer'. The $scale~value$ must be known, either from instrument characteristics or from comparison with other instruments or other data, for the data to be of use. 

The first Director of the observatory \citet{ner50} gave the scale value for the H-variometer as 3.6 nT/mm. The H-variometer is sensitive to temperature changes and the temperature was recorded, but the temperature coefficient was only determined later and the data then corrected for temperature variations. The scale value for the D-variometer calculated from the characteristics of the instrument is quoted as 0\farcm{315}/mm. No absolute measurements are given and the scale values are assumed to be constant, as there is no known meta-data about changes of calibration, which, however, does not necessarily mean that there were no such changes. We shall show here that significant changes took place and that the data need to be corrected accordingly, preferably from [as yet] undiscovered meta-data, if possible, and if not possible, from comparison with other observatories and other solar or geomagnetic proxies, e.g. as in \citet{mor14}.

\subsection{Calculating IDV(1d) from $s$ at ESK and HLS}
\begin{figure*}[t]
\vspace*{2mm}
\begin{center}
\includegraphics[width=18cm]{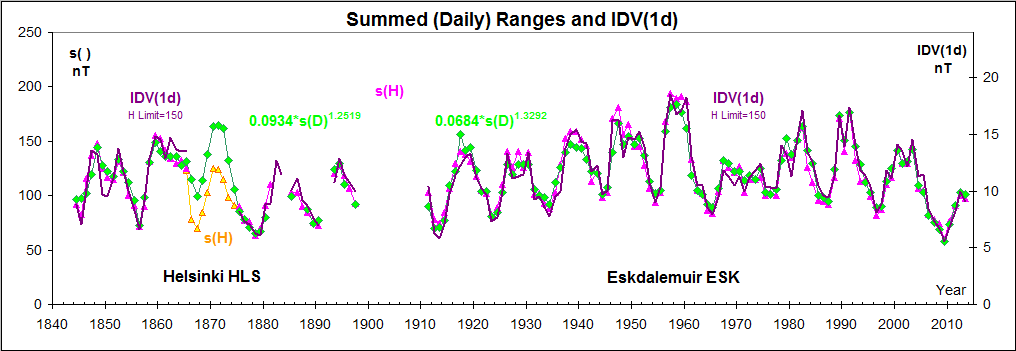}
\end{center}
\caption{Yearly average Summed Ranges for H (pink triangles) and for D (green circles) scaled to match the scale of H using the equations in green for HLS (left) and ESK (right). The equations are slightly different because the inhomogeneous `raw' values are plotted, i.e. not normalized to a common `bridge'. The values of s(H) for the interval 1866-1874.5 (orange symbols) do not match the rest of the s(H) to scaled s(D). IDV(1d) calculated from H for ESK (purple symbols; scale on right) is a good fit to s(H) and s(D). A few `spikes' have been suppressed by capping daily IDV(1d) at 150 nT. IDV(1d) calculated from H for HLS (scale at right; same scale as for ESK as no normalization between HLS and ESK is performed - this is raw data - for this inhomogeneous data set) is also a good fit to s(H) and s(D), except for the interval 1866-1874.5.}
\label{summed-ranges-IDV1d}
\end{figure*}

In \citet{lea13a,lea13b} the IDV(1d) series for HLS and ESK (Eskdalemuir) are spliced together using POT (Potsdam) as a 'bridge'. In spite of the bridge being at considerably lower corrected geomagnetic latitude (by 6\degree), it is posited that the result is a homogeneous data set. If so, results from ESK should be applicable to HLS as well. In Figure~\ref{summed-ranges-IDV1d} we show in the right-hand part the very similar variations since 1911 of the summed ranges s(H) and of s(D) [scaled to s(H)] and of IDV(1d) for ESK. This is as expected from the results demonstrated in section \ref{Summed-Ranges}. For HLS, shown in the left-hand part of the Figure, s(H) and [scaled] s(D) also agree closely, \textbf{except for the interval 1866-1874.5} [orange data points]. The simplest, and in our view inescapable, conclusion to draw from this discrepancy is that the adopted scale value of the H-variometer was too small, by some 30\%, during 1866-1874.5. Figure \ref{IDV1d-HLS} in the Supplemental Information [SI] compares IDV(1d) for HLS calculated from the summed ranges, further visualizing the obvious discrepancy. IDV(1d) for HLS, calculated for years outside of the interval 1866-1874.5  is plotted in Figure \ref{summed-ranges-IDV1d} as well, for comparison. The agreement with s(H) and s(D) is again good.

\section{IHV also Shows the Scale Value Discrepancies}
\begin{figure*}[h]
\vspace*{2mm}
\begin{center}
\includegraphics[width=8.3cm]{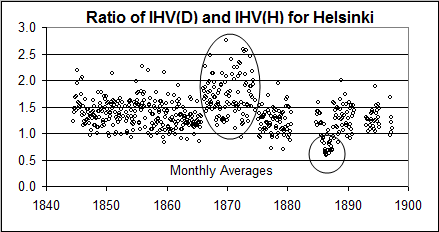}
\end{center}
\caption{(The ratio between monthly values of IHV calculated using the Declination, IHV(D), and of IHV calculated using the Horizontal Force, IHV(H) for Helsinki. The ovals show the effect of the scale value for H being too low 1866-1874.5 and of the scale value for D being too low 1885.8-1887.5.}
\label{IHV-HLS}
\end{figure*}

Svalgaard and colleagues \citep{sea03,sea04,sva07} introduced the InterHourly Variability index, IHV, as a proxy for auroral zone activity [as measured at mid-latitudes]. Although HLS is too close to the auroral zone for IHV calculated from HLS data to retain its simple physical meaning [a proxy for solar wind $BV^2$ and for the NOAA/POES hemispheric power \citep{eme08}], the IHV values do depend directly on the scale value used for the variometers. As for IDV, IHV can be computed for any geomagnetic element. If the scale values for H and for D were both correct, the ratio between IHV(D) and IHV(H) would be constant (apart from a random noise component). Figure \ref{IHV-HLS} shows the ratio between IHV(D) and IHV(H) for HLS. As predicted from the analysis in Section \ref{Scale-Values} the ratio shows the expected behavior (in large oval) for the interval 1866-1874.5. The smaller oval shows that there is also a problem with the scale value of the D-variometer in 1886-1887.5, being too low by about a factor of 2. This is explored further in the SI [Figure \ref{D-Problem}].

\section{The Daily Variation}
\begin{figure*}[t]
\vspace*{2mm}
\begin{center}
\includegraphics[width=18cm]{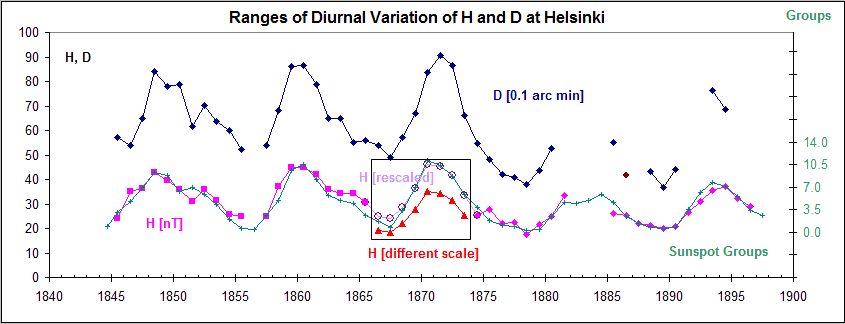}
\end{center}
\caption{Yearly average Ranges for Declination D [in 0.1 arc minute units], blue curve, and for Horizontal Force [in nT units], pink curve. Because of the strong seasonal variation only years with no more than a third of the data missing are plotted. The green curve [with ‘+’ symbols] shows the number of active regions [sunspot groups] on the disk scaled to match the pink curves (H). As expected the match is excellent, except for the interval 1866-1874, where the H-range would have to be multiplied by 1.32 for a match: purple open circles.}
\label{Range-SR-HLS}
\end{figure*}

The diurnal variation [$S_{R}$ sometimes less accurately called $Sq$] of geomagnetic elements can be used to check the scale value of the magnetometers. Computing for each day the differences between the instantaneous hourly values [or hourly means – the distortion caused by averaging over an hour is but slight] and the daily mean removes the effect of the (slowly varying) secular values and of random (unknown) changes in the baseline. The average, over an interval – such as a month or a year, of the differences as a function of time within the day is the average diurnal variation [what used to be called the daily ‘inequality’]. It is well-known that that average range, i.e. the difference between the maximum and minimum values of the average diurnal variation is extremely well correlated with appropriate solar activity indices (e.g. F10.7 microwave flux, sunspot number, or the group number (number of active regions on the solar disk)), as was discovered by \citet{wol52}\footnote{"Wer h\"atte noch vor wenigen Jahren an die M\"oglichkeit gedacht, aus den Sonnenfleckenbeobachtungen ein terrestrisches Ph\"anomen zu berechnen?"} and subsequently extensively verified by many workers \citep[e.g.][]{bar46}, considered to be the best of all solar-terrestrial correlations; a fact used by \citet{nea92} who note ``(t)he scale value for the D-variometer seems to be reliable (for 1844-1853) because the diurnal variation at the Helsinki, Nurmij\"arvi and St. Petersburg observatories show very similar behavior being the same within 1' under corresponding solar activity conditions".  Figure \ref{Range-SR-HLS} shows the yearly average ranges for Declination D and Horizontal Force H at Helsinki.

The Group numbers used in Figure \ref{Range-SR-HLS} are derived from the recent re-evaluation of solar activity \citep{sva13}. Using the official SIDC sunspot number does not change the result for the interval of interest (SI Figures  \ref{H-Problem} and  \ref{D-Problem}) . It seems that the scale value adopted for H during the interval 1866-1874 must actually be different from that used for the rest of the H-data, specifically 1.32 times lower than the constant value used by Nevanlinna in constructing the Helsinki series. The range of the Declination during that interval matches that of H when H is re-scaled upwards by the factor 1.32. The ranges of D and H generally vary together (with solar activity) being due to the same current system, so the discrepancy indicates a problem with the adopted scale-value of H. Figure \ref{H-Problem} in the SI documents the problem for the H-component for the year 1869, while Figure \ref{D-Problem} in the SI documents the problem for the Declination for the year 1886.

An equally strong case can also be made comparing the diurnal range of H at Helsinki directly to that of H at other stations, not using the solar activity connection. In Figure \ref{Range-Stations} of the SI, we show a comparison with Greenwich (GRW, brown), Prague (PRA, blue), and Colaba (CLA and replacement station Alibag ABG, green). Because not all stations observed hourly values all the time, the ranges have been matched to Helsinki (HLS, pink) outside the interval 1866-1874. During that interval, the range for HLS (red triangles) is seriously too low. 
 
\section{Reconstructed HMF}
\begin{figure*}[h]
\vspace*{2mm}
\begin{center}
\includegraphics[width=18cm]{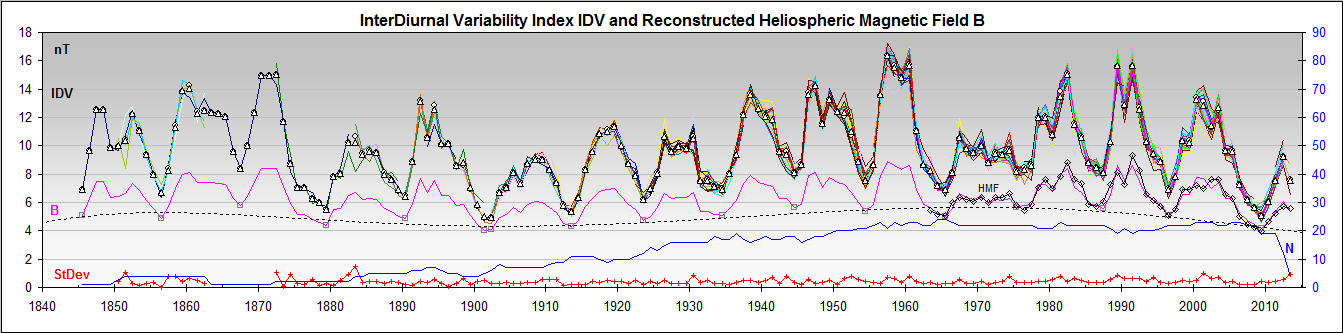}
\end{center}
\caption{Reconstruction of annual means of 169 years of near-Earth Heliospheric Magnetic Field strength B (pink line in middle of graph) 1845-2013 compared with in-situ spacecraft measurements (black line marked HMF) plotted using different colors for each station, from \citet{sva14}. Open triangles (or circles) show the median (or mean) of all stations in each year. The red line at the bottom of the graph shows the standard deviation of the values of IDV in each year. The blue line marked N shows the number of stations for each year.}
\label{HMF-B13}
\end{figure*}

Figure \ref{HMF-B13} shows a reconstruction of annual means of 169 years of near-Earth Heliospheric Magnetic Field strength $B$ (pink line in middle of graph) 1845-2013 compared with in-situ spacecraft measurements (black line marked HMF). The reconstruction is based on a re-evaluation of the IDV-index using the normalized average of the three determinations discussed in sections \ref{Summed-Ranges} and \ref{Scale-Values}; from \citet{sva14}, plotted using different colors for each station. For 1863-1871 leading up to the strong solar cycle 11, only HLS (Helsinki) contributes (awaiting digitization of other stations), underscoring the importance of getting HLS right.

A consequence of the undue weight given to the regular diurnal variation that we referred to in section \ref{Summed-Ranges} when using the Summed Ranges based on monthly averages that Bartels used to extend the u-measure before 1872 is that our earlier reconstruction based on the u-measure \citep{sva10} for years with large coronal holes and ensuing large HMF during the declining phase of the solar cycles was too low for years during the declining phase. Going to the Summed \textit{Daily} Ranges, s(H,D), remedies that defect, especially when the erroneus scale values are corrected. It is instructive to compare the reconstruction for solar cycles 10-11 (1857-1878) with those of cycles 18-19 (1945-1964), as regard to both the `shape' of the solar cycle curves and to the similar general level of activity.

\begin{figure*}[h]
\vspace*{2mm}
\begin{center}
\includegraphics[width=16cm]{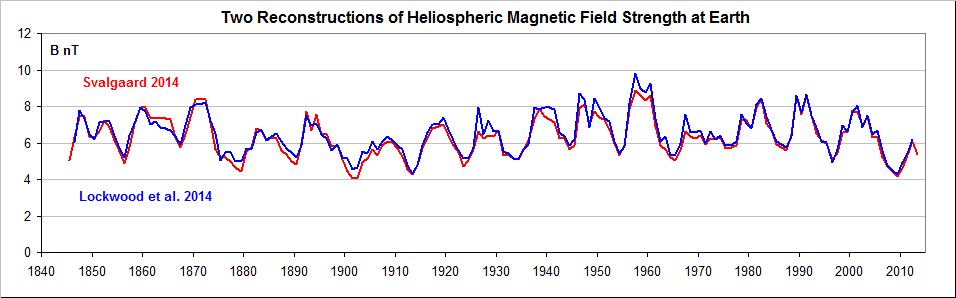}
\end{center}
\caption{Comparison of the HMF strength deduced by \citet{sva14} as shown in Figure \ref{HMF-B13} [red curve] and inferred by \citet{lea14a} [blue curve]. The coefficient of determination is $R^2= 0.93$.}
\label{Comparison}
\end{figure*}

After the present paper was submitted, Lockwood et al. \citep{lea14a,lea14b} and \citet{loc14}, now aware of our finding, have accepted our analysis and corrected their reconstruction accordingly. Figure \ref{Comparison} shows that their revised values \citep{lea14a} are largely correct (compared with our multi-station reconstruction), and that reconstruction of the HMF strength is now satisfactory constrained back to 1845. A significant insight that follows from the concordant reconstructions is that there hardly was any `Modern Grand Maximum' as the values of the HMF in the 20th century are on par with the values in the mid-19th.

\conclusions[Conclusion and Recommendations]  
Using several lines of evidence we have shown that the scale-values for the Helsinki magnetic data are in error at times. For H, the scale-value for 1866-1874 is too low by $\sim$30\% and for D too low during 1886 by a factor $\sim$2. \citet{loc13} reminds us about ``the great importance of knowing, as far as is possible, the true provenance of historic data and of all the corrections and changes that may have subsequently been applied to them". This is of critical importance in this situation where there is but one station for several years with usable data, so we urge Nevanlinna et al. to continue to re-examine the original data and their reduction. And we urged Lockwood et al. to revise accordingly their analysis and derivation of IDV look-alikes where based on the Helsinki data. Such correction and revision has now happened: in \citet{lea14b} they write ``a correction to seven years' IDV(1d) data during solar cycle 11 was discussed, checked against newly-analysed independent data from St Petersburg and implemented". This shows the value and power of comparisons with other data by the authors themselves and independently by other researchers.

\begin{acknowledgements}
We acknowledge the impetus from participating in work of the ISSI International Team 233 and from serving as reviewers of \citet{lea13a,lea13b}. LS appreciates the continuing support from Stanford University.
\end{acknowledgements}



\onecolumn

\clearpage


\begin{figure*}[t]
\vspace*{2mm}
\large{\textbf{Supplementary Information}\\In this section we collect various Figures providing supplementary support for the analysis in the paper.}
\label{}
\end{figure*}

\begin{figure*}[b]
\vspace*{2mm}
\begin{center}
\includegraphics[width=8.3cm]{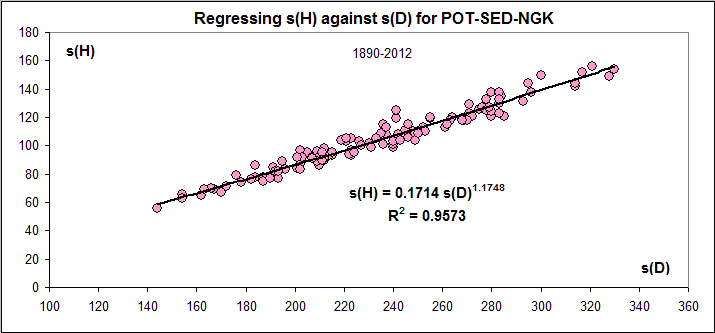}
\end{center}
\caption{The average s(H) for each year is plotted against the average s(D) for that year. The data can be fitted to a power law as shown which ‘explains’ 96\% of the correlation. We use power laws because most regression plots show somewhat curved point clouds [‘rivers’ is probably a more descriptive term].}
\label{Regression-sH-sD}
\end{figure*}

\begin{figure*}[b]
\vspace*{2mm}
\begin{center}
\includegraphics[width=8.3cm]{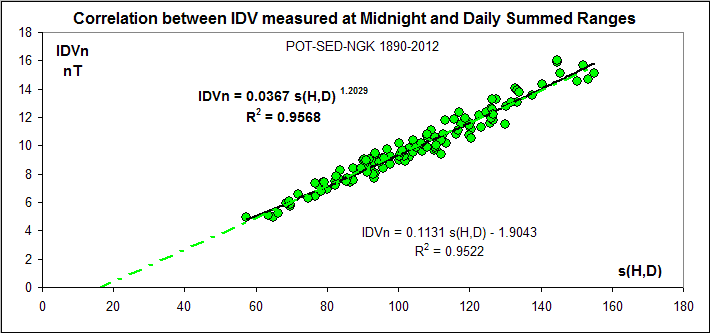}
\end{center}
\caption{Correlation between yearly values of IDVn [midnight] and the Average Summed Ranges for the day for H and D, s(H,D), for the POT-SED-NGK composite series 1890-2012. The dashed line is the linear relation extrapolated to vanishing IDVn.}
\label{Regression-IDV-s}
\end{figure*}

\begin{figure*}[t]
\vspace*{2mm}
\begin{center}
\includegraphics[width=16.6cm]{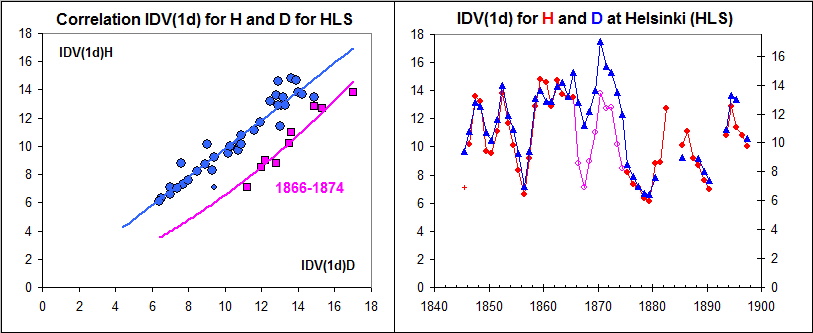}
\end{center}
\caption{Using s(D) and s(H) we can calculate the corresponding IDV(1d) values for HLS: right-hand panel. There is generally a good agreement between values derived derived using s(H) and s(D), except for the interval 1866-1874 for s(H) as also shown in the left-hand panel, plotting IDV(1d) derived from s(H) against IDV(1d) derived from s(D). Pink symbols are for 1866-1874 while blue symbols are for data outside of that interval. }
\label{IDV1d-HLS}
\end{figure*}

\begin{figure*}[t]
\vspace*{2mm}
\begin{center}
\includegraphics[width=16.6cm]{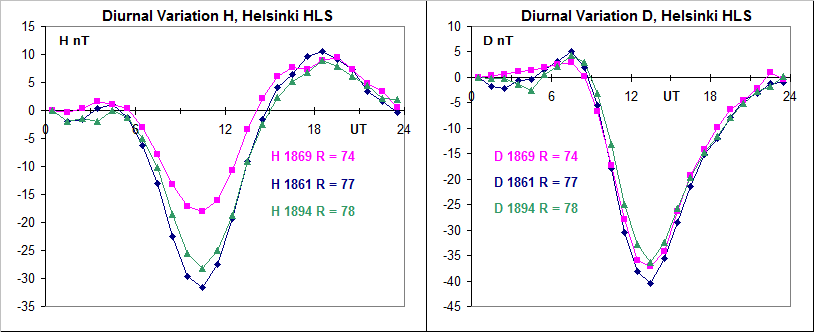}
\end{center}
\caption{Diurnal variation of H [left] and of D [right] at Helsinki for three years with SIDC sunspot number $\sim$75. It is very hard to escape the conclusion that the range of H for the year 1869 is too small compared to the other years with similar sunspot number as the range of D is about the same for the three years shown. The timing from 1882 on has been adjusted to 1 hour later (also in the following Figure) because of a change from G\"ottingen time to local Helsinki time which is not reflected correctly in the published data.}
\label{H-Problem}
\end{figure*}

\begin{figure*}[t]
\vspace*{2mm}
\begin{center}
\includegraphics[width=16.6cm]{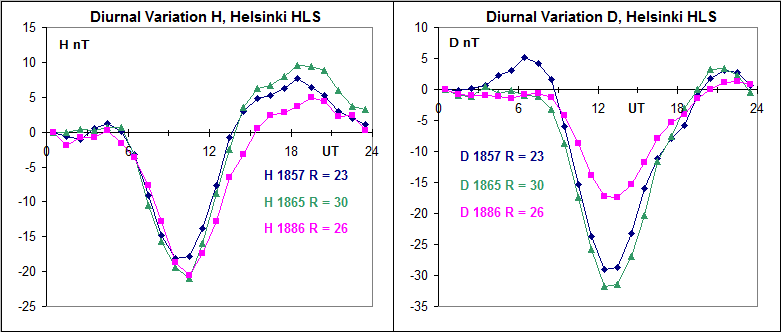}
\end{center}
\caption{Diurnal variation of H [left] and of D [right] at Helsinki for three years with SIDC sunspot number $\sim$25. It is very hard to escape the conclusion that the range of D for the year 1886 is too small compared to the other years with similar sunspot number as the range of H is about the same for the three years shown. Detailed analysis shows that the problem exists from October 1885 through May 1887. In August 2003 we emailed Nevanlinna alerting him to this problem, but, unfortunately, no corrective action resulted from this. It is now clear that scale-value problems exist for both the H component and for the Declination and that corrective action is mandatory before use of the Helsinki data.}
\label{D-Problem}
\end{figure*}

\begin{figure*}[t]
\vspace*{2mm}
\begin{center}
\includegraphics[width=16.6cm]{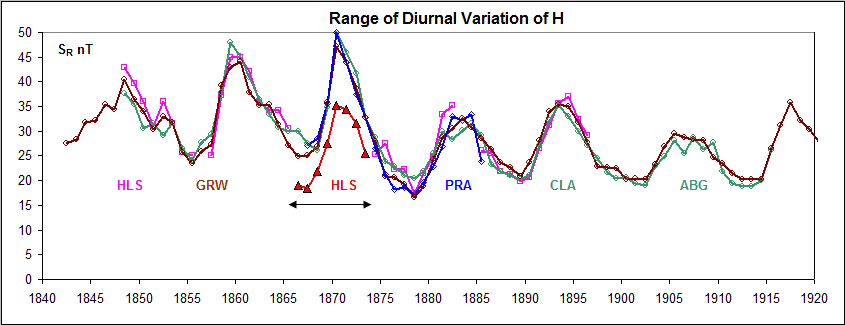}
\end{center}
\caption{The diurnal range (in nT) of the Horizontal Force for Prague (PRA) blue, Colaba (CLA)+Alibag (ABG) green, Greenwich (GRW) brown, and Helsinki (HLS) pink. For 1866-1874, HLS (red triangles) is clearly seriously too low. }
\label{Range-Stations}
\end{figure*}


\begin{thebibliography}{}

\bibitem[Bartels(1925)]{bar25}
Bartels, J.: Archiv des Erdmagnetismus, Heft 5, Abh. Met. Inst., 8(2), Berlin, 1925.  

\bibitem[Bartels(1932)]{bar32}
Bartels, J.: Terrestrial-magnetic activity and its relations to solar phenomena,  Terr. Magn. Atmos. Elec., 37, 1-52, 1932.  

\bibitem[Bartels(1946)]{bar46}
Bartels, J.: Geomagnetic data on variations of solar radiation, 1, Wave-radiation, Terr. Magn. Atmos. Elec., 51, 181-242, 1946.  

\bibitem[Emery et al.(2008)]{eme08}
Emery, B. A., Coumans, V., Evans, D. S., Germany, G. A., Greer, M. S., Holeman, E., Kadinsky-Cade, K., Rich, F. J., Xu, W.: Seasonal, Kp, solar wind, and solar flux variations in long-term single-pass satellite estimates of electron and ion auroral hemispheric power, J. Geophys. Res.,113(A6), A06311, 2008.

\bibitem[Lockwood et al.(1999)]{lea99}
Lockwood, M., Stamper, R., and Wild, M. N.: A doubling of the Sun's coronal magnetic field during the past 100 years, Nature, 399, 437, doi:10.1038/20867, 1999. 

\bibitem[Lockwood(2013)]{loc13}
Lockwood, M.: Reconstruction and Prediction of Variations in the Open Solar Magnetic Flux and Interplanetary Conditions, Living Rev. Solar Phys. 10, 4, doi:10.12942/lrsp-2013-4, 2013. \url{http://www.livingreviews.org/lrsp-2013-4}.

\bibitem[Lockwood \& Owens(2011)]{loc11}
Lockwood, M. and Owens, M. J.: Centennial changes in the heliospheric magnetic field and open solar flux: The consensus view from geomagnetic data and cosmogenic isotopes and its implications, J. Geophys. Res., 116(A4), A04109, doi:10.1029/2010JA016220, 2011.

\bibitem[Lockwood \& Owens(2014)]{loc14}
Lockwood, M. and Owens, M. J.: Implications of the Recent Low Solar Minimum for the Solar Wind During the Maunder Minimum, ApJL, 781:L7, doi:10.1088/2041-8205/781/1/L7, 2014.

\bibitem[Lockwood et al.(2013a)]{lea13a}
Lockwood, M., Barnard, L., Nevanlinna, H., Owens, M. J., Harrison, R. G., Rouillard, A. P., and Davis, C. J.:
Reconstruction of Geomagnetic Activity and Near-Earth Interplanetary Conditions over the Past 167 Years: 1. A New Data Composite,  Ann. Geophys., 31, 1957-1977, 2013a.

\bibitem[Lockwood et al.(2013b)]{lea13b}
Lockwood, M., Barnard, L., Nevanlinna, H., Owens, M. J., Harrison, R. G., Rouillard, A. P., and Davis, C. J.:
Reconstruction of Geomagnetic Activity and Near-Earth Interplanetary Conditions over the Past 167 Years: 2. A New Reconstruction of the Interplanetary Magnetic Field,  Ann. Geophys., 31, 1979-1992, 2013b.

\bibitem[Lockwood et al.(2014a)]{lea14a}
Lockwood, M., Barnard, L., Nevanlinna, H., Owens, M. J., Harrison, R. G., Rouillard, A. P., and Davis, C. J.:
Reconstruction of Geomagnetic Activity and Near-Earth Interplanetary Conditions over the Past 167 Years: 3. Improved 
representation of solar cycle 11,  Ann. Geophys., (in press), 2014a.

\bibitem[Lockwood et al.(2014b)]{lea14b}
Lockwood, M., Barnard, L., Nevanlinna, H., Owens, M. J., Harrison, R. G., Rouillard, A. P., and Davis, C. J.:
Reconstruction of Geomagnetic Activity and Near-Earth Interplanetary Conditions over the Past 167 Years: 4. Near-
Earth Solar Wind Speed, IMF, and Open Solar Flux,  Ann. Geophys., (in press), 2014b.

\bibitem[Mayaud(1980)]{may80}
Mayaud, P. N.: Derivation, Meaning, and Use of Geomagnetic Indices, Geophys. Monogr. Ser., vol. 22, pp. 154, ISBN 0-87590-022-4, American Geophysical Union, Washington D. C., 1980.

\bibitem[Morozova et al.(2014)]{mor14}
Morozova, A. L., Ribeiro, P., and Pais, M. A.: Correction of artificial jumps in the historical geomagnetic measurements of Coimbra Observatory, Portugal, Ann. Geophys., 32, 19-40, 2014.

\bibitem[Moos(1910)]{moo10}
Moos, N. A. F.:  Colaba Magnetic Data, 1846 to 1905, 2, The Phenomenon and its 
Discussion, 782 pp., Central Govt. Press, Bombay, 1910.

\bibitem[Nervander(1850)]{ner50}
Nervander, J. J.: Observations faites \`a l'observatoire Magn\'etique et M\'et\'eorologique de Helsingfors 1844-1848, Vol. I-IV, Helsingfors, 1850.

\bibitem[Nevanlinna et al.(1992)]{nea92}
Nevanlinna, H., Ketola, A., and Kangas, T.: Magnetic Results from Helsinki Magnetic-Meteorological Observatory, Finn. Met. Inst., Geophys. Publ., 27, pp. 155, 1992.

\bibitem[Nevanlinna(2004)]{nev04}
Nevanlinna, H.: Results of the Helsinki magnetic observatory 1844–1912, Ann. Geophys., 22, 1691–1704, 2004.

\bibitem[Svalgaard(2013)]{sva13}
Svalgaard, L.: Building a Sunspot Group Number Backbone Series, \url{http://www.leif.org/research/SSN/Svalgaard11.pdf}, 2013.

\bibitem[Svalgaard \& Cliver(2003)]{sva03}
Svalgaard, L. and Cliver, E. W.: New Geomagnetic Index (IDV) Measuring Magnitude of Interplanetary Magnetic Field, AGU Fall SH21B-0108,  \url{http://www.leif.org/research/AGU%20Fall%202003%20SH21B-0108.pdf, 2003.}

\bibitem[Svalgaard et al.(2003)]{sea03}
Svalgaard, L., Cliver, E. W., and Le Sager, P.: Determination of interplanetary magnetic field strength, solar wind speed, and EUV irradiance, 1890-2003, in Proceedings of ISCS 2003 Symposium: Solar Variability as an Input to the Earth's Environment, Eur. Space Agency Spec. Publ., ESA SP-535, 15, 2003. 

\bibitem[Svalgaard et al.(2004)]{sea04}
Svalgaard, L., Cliver, E. W., and Le Sager, P.: IHV: A new long-term geomagnetic index, Adv. Space. Res., 34, 436, 2004.
 
\bibitem[Svalgaard \& Cliver(2005)]{sva05}
Svalgaard, L. and Cliver, E. W.: The IDV index: Its derivation and use in inferring long-term variations of the
interplanetary magnetic field strength, J. Geophys. Res., 110(A12), A12103, doi:10.1029/2005JA011203, 2005.

\bibitem[Svalgaard \& Cliver(2007)]{sva07}
Svalgaard, L. and Cliver, E. W.: Interhourly variability index of geomagnetic activity and its use in deriving the
long-term variation of solar wind speed, J. Geophys. Res., 112(A10), A10111, doi:10.1029/2007JA012437, 2007.

\bibitem[Svalgaard \& Cliver(2010)]{sva10}
Svalgaard, L. and Cliver, E. W.: Heliospheric magnetic field 1835-2009, J. Geophys. Res., 115(A9), A09111,\\
doi:10.1029/2009JA015069, 2010.

\bibitem[Svalgaard \& Cliver(2014)]{sva14}
Svalgaard, L. and Cliver, E. W.: Update: Heliospheric magnetic field 1835-2013, J. Geophys. Res., (paper in preparation), 2014.

\bibitem[Wolf(1852)]{wol52}
Wolf, J. R.: Neue Untersuchungen \"uber die Periode der Sonnenflecken und ihre Bedeutung, Viertel. Natur. Ges. Bern, 245, 249-270, Bern, 1852.

\end{thebibliography}
\end{document}